\documentstyle[preprint,tighten,aps]{revtex}
 
\begin{document} 
\draft 
\preprint{
\begin{tabular}{r} FTUV/98$-$10 \\ IFIC/98$-$10
\end{tabular}
}
\title{Electric polarizability of nuclei from a longitudinal sum rule}
\author{J.\ Bernab\'eu$^a$, D.\ G\'omez Dumm$^a$, G.\ Orlandini$^b$}
\address{\hfill \\
$^a$ Departament de F\'{\i}sica Te\`orica, Universitat de Val\`encia, \\
Dr.\ Moliner 50, E-46100 Burjassot (Val\`encia), Spain\\
$^b$  Dipartimento di Fisica, Universit\`a di Trento,
I-38050 Povo, Italy \\and Istituto Nazionale di Fisica Nucleare, 
Gruppo collegato di Trento}
\maketitle  
\thispagestyle{empty}

\begin{abstract}
The nuclear electric polarizability is theoretically analyzed using a
sum rule derived from the longitudinal part of the forward Compton
amplitude. Beyond the leading dipole contribution, this approach leads
to the presence of potential-dependent terms that do not show up in
previous analyses. The significance of these new contributions is
illustrated by performing an explicit calculation for a proton-neutron
system interacting via a separable potential.
\end{abstract}

\hfill


\pacs{ }

\setcounter{page}{2} 

\section{Introduction} 

For many years, the study of the electric and magnetic 
polarizabilities of hadronic systems has deserved a significant 
interest towards the understanding of strong interactions. 
The polarizabilities account for the distortion of a system (induced 
dipole moments) in the presence of external quasistatic electric 
and magnetic fields, so that they determine the lowest-order 
response of the system's internal degrees of freedom to the
electromagnetic interactions \cite{review}. In particular, in the case of
a real Compton scattering (RCS) process, it is well known from low-energy
theorems that the corresponding spin-averaged forward scattering amplitude
is determined up to second order in the photon energy by the sum of both
the electric and magnetic polarizabilities, $\alpha+\beta$ \cite{low}.
One has indeed
\begin{equation} 
f(\omega^2)= -\frac{Z^2 e^2}{4\pi M} + (\alpha+\beta)\, \omega^2 +
O (\omega^4) \,,
\label{fw} 
\end{equation} 
where $\omega$ is the energy of the (real) photon, and the first term on
the right hand side corresponds to the Thomson limit, which depends only
on global properties of the system: its mass and its electric charge.

The use of a forward dispersion relation allows to find a connection
between the sum $\alpha+\beta$ in (\ref{fw}) and the total
photo-absorption cross section for the system, $\sigma_{tot}(\omega)$.
This relation can be expressed in the form of a sum rule \cite{suma},
\begin{equation} 
\alpha+\beta=\frac{1}{2\pi^2}\int^\infty_{\omega_{th}} 
d\omega\;\frac{\sigma_{tot}(\omega)}{\omega^2} \,,
\end{equation}
where $\omega_{th}$ stands for the threshold inelastic excitation of the
system. Thus the sum of the structure-dependent polarizabilities can be
estimated by performing experimental measurements of $\sigma_{tot}
(\omega)$ up to large enough photon energies.

Unfortunately, this approach cannot be used to determine both
parameters $\alpha$ and $\beta$ independently, as it is clear from the fact
that only the sum $\alpha+\beta$ appears in (\ref{fw}). In order to get
separate sum rules from RCS processes, it is necessary to take into account
also nonforward amplitudes, which in general lead to the introduction of
additional difficulties. This is e.g.\ the case if one considers the
backward amplitude, which is related to the difference $\alpha-\beta$; it can
be seen that the corresponding dispersion relation needs
information not only on the s-channel photo-absorption cross sections, but
also on the t-channel two-photon processes \cite{bef}. As a consequence,
the analysis turns out to be more involved and includes some model dependence.

There is, however, an alternative possibility. As it was
shown in Ref.~\cite{bt} some time ago, it is possible to obtain a sum rule
for $\alpha$ that involves the longitudinal part of the {\em virtual} photon
cross section $\sigma_L(\omega,q^2)$ in the region of quasi-real photons.
In fact, although the longitudinal cross section itself vanishes in the
on-shell photon limit, it is found that the integral of the slope of
$\sigma_L(\omega,q^2)$ at $q^2=0$ is directly related to the electric
polarizability. The sum rule can be
derived by considering an unsubtracted dispersion relation for the
longitudinal forward virtual Compton amplitude $T_L(\omega,q^2)$, with the
only ingredients of gauge invariance, analiticity and unitarity. The result
in Ref.~\cite{bt} explicitly reads
\begin{equation} 
\alpha - (\frac{e^2}{4\pi}) \frac{\mu^2}{4 M^3} = 
\frac{1}{2\pi^2}\int^\infty_{\omega_{th}} d\omega\; 
\lim_{q^2\rightarrow 0} \frac{\sigma_L(\omega,q^2)}{-q^2} \,,
\label{regla} 
\end{equation} 
where $\mu$ is the corresponding anomalous magnetic moment. We stress 
that the presence or not of a subtraction for $T_L$ cannot be inferred 
from the Thomson limit, therefore Eq.\ (\ref{regla}) 
is valid under the assumption that the integrand in the right hand side
has an adequate high-energy behaviour.

In this paper, we make use of the above longitudinal sum rule to derive
a theoretical expression for the electric polarizability of nuclear
systems. In order to deal with the nuclear photo-absorption cross section,
we proceed by performing a conventional nonrelativistic treatment of the
nuclear interactions, keeping only leading terms in powers of the inverse
mass of the system. In this regard, however, there is an essential
difference with respect to previous approaches: since our starting point
for the evaluation of the electric polarizability is relation
(\ref{regla}), the nonrelativistic approximations are introduced here only
{\em after} having taken into account the properties of gauge invariance,
analyticity and unitarity of the $S$ matrix, which have already been
invoked to obtain the sum rule. In contrast, these properties are not
exactly satisfied from the very beginning when the nonrelativistic limit
is taken directly on the real part of the Compton amplitude. This
last procedure has been followed e.g.\ in Refs.~\cite{otros,friar}.
We find that our treatment allows to improve the result for $\alpha$ obtained
in these previous analyses, leading to the presence of additional
contributions that depend on the nuclear interaction.
To evaluate the significance of these contributions, we
consider the simple case of a proton-neutron system interacting via a
separable Yamaguchi potential.

The calculation of the electric polarizability from the sum rule
(\ref{regla}) is performed in Section II. In Section III we present
the application to the Yamaguchi-like interacting system, while
Section IV contains our conclusions.
 
\section{Nuclear electric polarizability} 

Starting from the sum rule in Eq.\ (\ref{regla}), we derive in this
section the leading contributions to the nuclear electric polarizability,
up to first order in the inverse nuclear mass $M_A^{-1}$. To this end, it
is convenient to write the integrand on the right hand side of
(\ref{regla}) in terms of the so-called longitudinal response, which is
measured in many nuclei by means of inelastic electron scattering.
This function is defined as
\begin{equation}
R_L(|{\bf q}|,\omega)=\sum_{n>0}|\langle n|\rho({\bf q})|0\rangle |^2
\,\delta(\omega-\omega_{rec}-E_n+E_0)\,,
\label{rl}
\end{equation}
where $\rho({\bf q})$ represents the charge operator, $\omega_{rec}=
|{\bf q}|^2/(2M_A)$ is the  recoil energy of the nuclear system, and $E_0$
($|0\rangle$) and $E_n$ ($|n\rangle$) are the eigenvalues (eigenstates) of
the nuclear Hamiltonian corresponding to the ground and excited states,
respectively. For a spinless nucleus, it is easy to see that
Eq.\ (\ref{regla}) can be written as
\begin{equation}
\alpha={1\over 2\pi}\int_{\omega_{th}}^\infty\,d\omega\,
\left.{R_L(|{\bf q}|,\omega)
\over \omega \,|{\bf q}|^2}\right|_{|{\bf q}|\to\omega}\,.
\label{alfrl}
\end{equation}

From the analysis of electron scattering, it is seen that the longitudinal
response displays a strong collectivity at small momenta, whereas for
large momenta a single particle character is observed. In particular,
it has been shown \cite{ELO} that the proton-neutron dynamical correlations
generate a high-energy tail that can be relevant for the Coulomb sum rule.
We notice that, although this effect could also be significant in the
quasi-real photon limit, the high-energy contributions to the longitudinal
sum rule for $\alpha$ will be in general suppressed in view of the inverse
powers of $\omega$ entering the integrand in Eq.\ (\ref{alfrl}).

Let us work out the right hand side of (\ref{alfrl}). We begin by considering
the charge operator in (\ref{rl}), which at lowest relativistic order
is given by
\begin{equation}
\rho({\bf q})=e\sum_{j=1}^Z\exp{(i {\bf q}\cdot {\bf r}_j)}\,,
\end{equation}
where $Z$ is the number of protons and ${\bf r}_j$ are their coordinates
with respect to the nuclear center of mass. Then, performing an expansion in
powers of $|{\bf q}|$, one has
\begin{eqnarray}
\alpha&=&{e^2\over 2\pi}\int_{\omega_{th}}^\infty\,{d\omega\over \omega}
\,\lim_{|{\bf q}|\to\omega} \left\{
\delta\left(\omega-{|{\bf q}|^2\over 2M_A }-E_n+E_0\right) \,\sum_{n>0}
\,\Biggl[\,|\langle 0|C_1|n\rangle |^2 \right.\Biggr. \nonumber \\
& & \left.\left. + \, {|{\bf q}|^2\over 4}
\left(|\langle 0|C_2|n\rangle |^2-{4\over 3}
\langle 0|C_1|n\rangle\langle n|C_3|0\rangle \right) + O(|{\bf q}|^4)
\,\right]\,\right\}\,,
\label{siete}
\end{eqnarray}
where $C_n\equiv\sum_{i=1}^Z z_i^n$. In this expression, it is worth to
notice the presence of the electric dipole operator $C_1$ at leading order
in $|{\bf q}|$, while higher multipoles are also relevant to the
next-to-leading order (their importance will be seen later). Taking
explicitly the limit in the right hand side, and performing the integration
in $\omega$, the expression for $\alpha$ reads
\begin{eqnarray}
\alpha & = & {e^2\over 2\pi} \sum_{n>0}
\left(1-{\epsilon_n\over M_A}\right)^{-1}
\Biggl[\epsilon_n^{-1} |\langle 0|C_1|n\rangle |^2 \Biggr. \nonumber \\
& & \Biggl. + {\epsilon_n \over 4}
\left(|\langle 0|C_2|n\rangle |^2-{4\over 3}
\langle 0|C_1|n\rangle \langle n|C_3|0\rangle\right) +
O(\epsilon_n^2)\Biggr] \,,
\end{eqnarray}
where $\epsilon_n=M_A\left(1-\sqrt{1-2\,(E_n-E_0)/M_A}\,\right)$. 
Finally, expanding up to first order in $(E_n-E_0)/M_A$, one gets
\begin{eqnarray}
\alpha & = & \frac{e^2}{2\pi} \sum_{n>0} \left[
\frac{|\langle 0|C_1|n\rangle|^2}{E_n-E_0} + \frac{1}{2M_A}
|\langle 0|C_1|n\rangle|^2 \right. \nonumber \\
& & + \frac{1}{4}(E_n-E_0) \left( |\langle 0|C_2|n\rangle|^2 -\frac{4}{3}
\langle 0|C_1|n\rangle \langle n|C_3|0\rangle \right) \nonumber \\
& & + \Biggl. \frac{3}{8M_A} (E_n-E_0)^2
\left(|\langle 0|C_2|n\rangle|^2 -\frac{4}{3}
\langle 0|C_1|n\rangle\langle n|C_3|0\rangle\right) +  
O(M_A^{-2})\Biggr]\,.
\label{alf1}
\end{eqnarray}

Eq.\ (\ref{alf1}) gives $\alpha$ in terms of different energy-weighted
sums involving nuclear matrix elements of the $C_n$ operators $(n=1,2,3)$.
It is also possible to express this result by means of sum
rules, i.e.\ of average values of commutators and anticommutators of the
$C_n$ operators and the total nuclear Hamiltonian \cite{OTrep}. In order to
do this, let us first rewrite the nuclear polarizability in terms
of the moments $m_p$ of the longitudinal response function, defined by
\begin{equation}
m_p(|{\bf q}|)\equiv \int_{0^+}^\infty d\tilde\omega \;
(\tilde\omega)^p\, R_L(|{\bf q}|,\omega) \,,
\end{equation} 
where $\tilde\omega=\omega-|{\bf q}|^2/(2 M_A)$. It can be easily verified
that the following relations between the derivatives of the moments and the
energy-weighted sums hold:
\begin{mathletters}
\label{mom}
\begin{eqnarray}
m'_{-1}(0)&\equiv&\left.\frac{dm_{-1}(|{\bf q}|)}{d|{\bf q}|^2}
\right|_{|{\bf q}|^2=0} =
\sum_{n> 0} \frac{|\langle 0|C_1|n\rangle|^2}{E_n-E_0} \\
m'_{0}(0)&\equiv&\left.\frac{dm_0(|{\bf q}|)}{d|{\bf q}|^2}
\right|_{|{\bf q}|^2=0} = \sum_n |\langle 0|C_1|n\rangle|^2
\label{momb}  \\
m''_{1}(0)&\equiv &\left.\frac{d^2m_1(|{\bf q}|)}{(d|{\bf q}|^2)^2}
\right|_{|{\bf q}|^2=0} 
=  {1\over2}\sum_n (E_n-E_0) \left(|\langle 0|C_2|n\rangle|^2
-\frac{4}{3} \langle 0|C_1|n\rangle\langle n|C_3|0\rangle \right)  \\
m''_{2}(0)&\equiv & \left.\frac{d^2m_2(|{\bf q}|)}{(d|{\bf q}|^2)^2}
\right|_{|{\bf q}|^2=0}  = 
{1\over 2}\sum_n (E_n-E_0)^2 \left( |\langle 0|C_2|n\rangle|^2
-\frac{4}{3} \langle 0|C_1|n\rangle\langle n|C_3|0\rangle \right)
\label{momd}
\end{eqnarray}
\end{mathletters}
\hspace{-.27cm}
thus $\alpha$ can be written as
\begin{equation}
\alpha=\left(\frac{e^2}{4\pi}\right) \left[ 2\, m'_{-1}(0) +
\frac{1}{M_A} m'_0(0) + m{''}_1(0)+\frac{3}{2M_A} m{''}_2(0) + 
O\left(M_A^{-2}\right)\right]\,.
\label{alfyam}
\end{equation}
Now, by using the closure property, it can be shown that the derivatives
of the nonnegative moments in Eqs.\ (\ref{mom}) satisfy the following sum
rules\cite{OTrep}:
\begin{mathletters}
\label{momvac}
\begin{eqnarray}
m'_{0}(0)&=& {1\over 2}\langle 0|\{C_1,C_1\}|0\rangle \\
m''_{1}(0)&=& {1\over 4}\langle 0|\left[C_2,[H,C_2]\right]|0\rangle 
-{1\over 3}\langle 0|\left[C_1,[H,C_3]\right]|0\rangle \label{mom1} \\
m''_{2}(0)&=& 
{1\over 4}\langle 0|\{[C_2,H],[H,C_2]\}|0\rangle
-\frac{1}{3} \langle 0|\{[C_1,H],[H,C_3]\}|0\rangle \, . \label{mom2}
\end{eqnarray}
\end{mathletters}
\hspace{-.27cm}
Moreover, some parts of the commutators and anticommutators can be
explicitly evaluated; in particular, the double commutators containing
the kinetic part $T$ of the Hamiltonian ($H=T+V$) give
rise to terms proportional to the proton radius, leading to
\begin{mathletters}
\label{conm}
\begin{eqnarray}
\langle 0|\left[C_2,[H,C_2]\right]|0\rangle & = & {4\over m}
\left({1\over 3}\langle 0|\sum_{i=1}^Z {\bf r}_i^2|0\rangle - {1\over A}
\langle 0|C_1\,C_1|0\rangle \right)+\langle 0|\left[C_2,[V,C_2]
\right]|0\rangle
\label{conm1} \\
\langle0|\left[C_1,[H,C_3]\right]|0\rangle & = & {1\over m}\left(1-
{Z\over A}\right) \langle0|\sum_{i=1}^Z {\bf r}_i^2|0\rangle +
\langle 0|\left[C_1,[V,C_3]\right]|0\rangle\,,
\label{conm2}
\end{eqnarray}
\end{mathletters}
\hspace{-.27cm}
where we have approximated $M_A\simeq m\,A$, being $m$ the nucleon mass.
By making use of relations (\ref{mom}-\ref{conm}), the nuclear electric
polarizability can finally be written as
\begin{eqnarray}
\alpha&=& \left( \frac{e^2}{4\pi} \right) \left[ 2 \sum_{n > 0} 
\frac{|\langle 0|C_1|n\rangle |^2}{E_n-E_0} + \frac{Z}{3 M_A} \langle 0| 
\sum_{i=1}^Z {\bf r}_i^2|0\rangle +\frac{1}{4} \langle 0|[C_2,[V,C_2]]|0
\rangle\right. -\frac{1}{3} \langle 0|[C_1,[V,C_3]]|0\rangle  \nonumber \\ 
& & \left. + \frac{3}{8 M_A}
\langle 0|\{[C_2,V],[V,C_2]\}|0\rangle  -\frac{1}{2 M_A} 
\langle 0|\{[C_1,V],[V,C_3]\}|0\rangle + O(M_A^{-2})
\right]\,.
\label{main}
\end{eqnarray}

Eq.\ (\ref{main}) represents the main result of this work. It is seen
that, beyond the leading dipole term, there is a contribution to $\alpha$
proportional to $\langle {\bf r}^2\rangle$, plus potential-dependent
terms of order $V$ and $V^2/M_A$. Let us remark that previous nonrelativistic
analyses \cite{otros,friar} lead only to the first two terms in
(\ref{main}). The significance of the potential-dependent contributions
obtained from our calculation will be illustrated in the next section,
where we consider the case of a simple solvable proton-neutron system.
The study of Eq.\ (\ref{main}) for realistic
nuclear potentials leading to exchange-current contributions will be the
subject of analysis in future work.

As a final comment, let us point out that the above result includes both
the contributions coming from collective nuclear effects and
those arising from the polarizability of the individual nucleons inside the
nucleus. In fact, it can be seen that the photoabsorption cross section
appears to be dominated by the nucleonic excitations for energies above
the pion mass. Below this threshold, there is a region
where the cross section still shows a volume character, which can be
described in terms of the interaction of the photon with quasi-deuteron
systems. Thus, for these energies, the contribution to the polarizability
can also be understood as a nucleonic one, being the nucleons modified by
the surrounding medium \cite{ERC}. Within our formulation, the identification
of the volume contributions is not trivial, since
the result in Eq.\ (\ref{main}) is obtained after performing an integration
over the whole spectrum. We recall, however, that the sum rule for $\alpha$
considered here weights the longitudinal response inversely with
$\omega$, hence the effects coming from the high energy region are
in general expected to be suppressed.

\section{Proton--neutron system with a Yamaguchi potential} 

As an application of the above result, let us 
evaluate the contributions to $\alpha$ for a simple case, 
namely a proton-neutron system interacting via a separable potential 
\begin{equation} 
V=\lambda\, |g\rangle \langle g|\;,\hspace{1cm} \lambda <  0 \,.
\end{equation} 
Denoting by $\psi_0({\bf p})$ and $\psi_{\bf k}({\bf p})$ the
bound state wave function and the scattering solution for this potential
respectively, the longitudinal structure function $R_L$ will be given by
\begin{equation} 
R_L(|{\bf q}|,\omega)= \int d^3k\;
\delta(\omega-\frac{{\bf k}^2}{2M}-\epsilon_B)\,
\left|\int d^3p\;\psi_{\bf k}({\bf p})\,\psi_0({\bf p}-{\bf q})\right|^2 \,,
\label{sl}
\end{equation} 
where $M$ is the total mass of the system and $\epsilon_B$ stands for the
binding energy corresponding to the $\psi_0$ state. We consider for
simplicity the particular case of a Yamaguchi potential \cite{yam}, 
\begin{equation} 
\langle {\bf p}|g\rangle =\frac{\sqrt{\beta}}{\pi}\frac{1}{p^2+\beta^2}\;, 
\hspace{1cm} \epsilon_B=\frac{\beta^2}{2M} \;,
\label{yam}
\end{equation} 
which allows to perform the integrals in (\ref{sl}) analytically. After
some algebra, it is found \cite{rosen} that the longitudinal structure
function can be written in terms of the variables $x\equiv
\sqrt{\omega/\epsilon_B-1}$ and $y\equiv |{\bf q}|/\beta$ as
\begin{equation} 
R_L(y,x)=\frac{16}{\pi \epsilon_B} \frac{x}{D^3} \left[ 
(1+x^2+y^2)^2+\frac{4}{3} x^2 y^2 - D\, F_0(y) \left( 1+\frac{y^2}{2} 
-\frac{y^4}{2(1+x^2)}\right) \right] \,,
\label{slxy}
\end{equation} 
where $D\equiv (1-x^2+y^2)^2+4x^2$, and $F_0(y)$ is the elastic form 
factor,
\begin{equation}
F_0(y)=\frac{1}{Z} \langle\psi_0|\rho({\bf q})|\psi_0\rangle= 
\left(1+\frac{y^2}{4}\right)^{-2} \,.
\end{equation} 
In the case under consideration, the integrals corresponding to the moments
in (\ref{mom}) are convergent, so that the terms in the right
hand side of (\ref{alfyam}) can be evaluated using the expression for $R_L$
in (\ref{slxy}). Considering as before the terms in $\alpha$ up to
order $M^{-1}$, we have
\begin{equation}
\alpha=\left(\frac{e^2}{4\pi}\right) \left[ \frac{7}{12 M \epsilon_B^2}
+ \frac{1}{2M^2\epsilon_B} - \frac{3}{16 M^2 \epsilon_B} +
\frac{9}{8 M^3} \right] \,,
\label{alfin}
\end{equation}
where the four terms correspond to those in the right hand side of Eq.\
(\ref{alfyam}), respectively. Now, it is instructive to compare this
expression for $\alpha$ with the main result in Eq.\ (\ref{main}),
in order to identify the contribution of the potential-dependent terms for
this simple case. From (\ref{mom2}), we see that the $V^2$ part in
(\ref{main}) corresponds to the $m{''}_2(0)$
contribution, which yields the $M^{-3}$ term in (\ref{alfin}). This is
consistent with the nonrelativistic approximation, which assumes
implicitly that the nuclear interactions are relatively ``weak'', and both
the potential and kinetic energies should be considered to be order $(1/M)$
\cite{friar}. Furthermore, it is found that the remaining
potential contributions to (\ref{main}) correspond to the $m{''}_1(0)$
term. In fact, in the $Z=1$ case the kinetic contributions cancel out,
and we end up with
\begin{equation}
\frac{1}{4} \langle 0|[C_2,[V,C_2]]|0\rangle 
-\frac{1}{3} \langle 0|[C_1,[V,C_3]]|0\rangle = m{''}_1(0)=
-\frac{3}{16M^2\epsilon_B}
\label{pot}
\end{equation}
Finally, we see that the contributions of $m'_{-1}(0)$ and $m'_0(0)$
(first two terms in (\ref{alfin})) correspond to the first two terms in
(\ref{main}), respectively. Then we conclude, at least for this model,
that {\em the lowest order potential-dependent contribution to $\alpha$
(i.e.\ that in (\ref{pot})) has the same order of magnitude as the
$\langle {\bf r}^2 \rangle$ term}. This is once again consistent with
treating the potential as order $1/M$. On the other hand, we recall that
previous nonrelativistic calculations for $\alpha$, such as those in
Refs.~\cite{otros,friar} led only to the $\langle{\bf r}^2\rangle$
correction. In this simple example we find that the additional
potential-dependent terms provided by the longitudinal sum rule approach
contribute significantly to the electric polarizability and therefore
should also be taken into account.

\section{Conclusions}

We have presented here a novel approach to analyze the electric
polarizability of nuclear systems. The distinctive feature of our analysis is
the use of the sum rule displayed in Eq.\ (\ref{regla}), which arises from
the assumption of an unsubtracted dispersion relation for the longitudinal
forward virtual Compton amplitude. This sum rule involves the virtual
photo-absorption cross section of the system in the region of quasi-real
photons.

We have calculated the leading terms contributing to the electric
polarizability for a spinless nucleus, up to first order in the inverse
nuclear mass. To do this, we have assumed that the photo-absorption
longitudinal cross section shows an adequate high-energy behaviour, so
that the treatment of the nucleus as a nonrelativistic system is consistent
with the use of the sum rule.

The main result of our calculation is shown in Eq.\ (\ref{main}). As expected,
it is found that the lowest order contribution to $\alpha$ is given by the
inverse energy-weighted sum of the strengths of inelastic dipole
excitations. Beyond this leading order, we obtain
a contribution proportional to $\langle{\bf r}^2\rangle$, plus
potential-dependent terms that appear
to be also significant. In particular, if the
kinetic and potential energies are both considered to be order $M^{-1}$
(which is consistent with the nonrelativistic approximation), the
terms proportional to $V$ in (\ref{main}) are shown to be of the same
order of magnitude as the $\langle{\bf r}^2\rangle$ one.
This is e.g.\ the case for a proton-neutron system interacting through
a separable Yamaguchi potential. As stated above, the presence of the
$\langle{\bf r}^2 \rangle$ contribution at the order $M^{-1}$ had been
derived many years ago by explicit nonrelativistic calculations of the
Compton amplitude for a bound system. The origin of the new
potential-dependent corrections in our analysis can be traced back to the
properties of gauge invariance, analiticity and unitarity of the $S$ matrix,
which are implicitly taken into account once the sum rule for $\alpha$ has
been invoked.

It is amazing that the requirement of completely general conditions on
the virtual Compton scattering amplitude automatically leads to the
presence of exchange current contributions to the nuclear polarizability.
The analysis of such contributions for realistic nuclear potentials
deserves a detailed study that will be the subject of future work.

\acknowledgements

G.\ O.\ would like to thank the Departament de F\'{\i}sica Te\`orica
of the University of Valencia for warm hospitality. D.\ G.\ D.\ has
been supported by a grant from the Commission
of the European Communities, under the TMR programme (Contract 
N$^\circ$ ERBFMBICT961548). This work has been funded by CICYT,
Spain, under Grant AEN-96/1718.

\end{document}